\documentstyle[prb,aps,epsf]{revtex}
\begin{document}
\draft
\title{Localizing periodicity in near-field images}
\author{P. Fraundorf}
\address{Physics \& Astronomy, U. Missouri-StL (63121), \\
Corporate Research, Monsanto (63167), \\
Physics, Washington U. (63130), \\
St. Louis, MO, USA}
\date{\today }
\maketitle

\begin{abstract}
We show that Bayesian inference, like that used in statistical mechanics,
can guide the systematic construction of Fourier dark-field methods for
localizing periodicity in near-field (e.g. scanning-tunneling and
electron-phase-contrast) images. For crystals in an aperiodic field, the
Fourier coefficient $Ze^{i\varphi }$ combines with a prior estimate for
background amplitude $B$ to predict background phase ($\beta $) values
distributed with a probability $p(\beta -\varphi \mid Z,\varphi ,B)$
inversely proportional to the amplitude $P$ of the signal of interest, when
this latter is treated as an unknown translation scaled to $B$. From
UMStL-CME-90f26pf.
\end{abstract}




\section{Introduction}

Near-field images (defined here as images of wave amplitude or phase at the
exit surface of a solid) with atomic (i.e. less than 2\AA ) resolution are
developments of the last quarter of this century. Transmission electron
microscopes capable of delivering phase-contrast images with continuous
transfer to spatial frequencies beyond 1/(2\AA ) have become available in
the last decade,\cite{Spence} and the first scanning tunneling\cite{Binnig1}
and atomic-force\cite{Binnig2} microscopes able to resolve atoms have been
created in this period as well. More recently, near-field visible-light
microscopes with resolutions much below the wavelength limit normally
associated with light microscopes, although not with atomic resolution, have
been described as well.\cite{Betzig}

As viewed from frequency space, near-field images with atomic resolution may
contain data of three basic types. The first type, which we refer to as the
``diffraction data'', are simply the data on lattice periodicity amplitudes
contained in image power spectra. All researchers accustomed to obtaining
data on lattice parameters or orientation from diffraction patterns have
experience with this information. Near-field images, sometimes under a more
restricted set of conditions, can also contain ``phase-information'' on the
phase-lag from one periodicity to the next. Diffractionists involved in
structure determination will recognize that such information is needed,
along with diffraction data, to determine the distribution of scattering
density within unit cells. Finally, such images also contain ``darkfield
information'', which we define here as information on amplitude and phase
differences across the breadth of individual diffracted beams. This
information tells how near-by periodicities interfere via ``beat''
processes, and hence how the intensity of any given range of periodicities
is distributed throughout the region examined. In other words, it tells
where crystals and their boundaries are located in the image field.
Researchers involved in small-angle and anomalous Bragg-scattering studies,
as well as in diffraction imaging techniques (like x-ray topography or
weak-beam electron imaging), all use this sort of information. Electron and
x-ray dark-field techniques which use far-field diffraction contrast to
provide data on both phase and amplitude components of this information, in
particular, have served for decades as powerful tools for the study of
defects in (and boundaries around) crystals.\cite{Hirsch}

Although the darkfield data in high-resolution near-field images are of
demonstrable interest for complementing that available via the far-field
techniques discussed above, methods for extracting that information have
generally taken the form of qualitative and {\em ad hoc} ``recipes'' for
image processing.\cite{Buseck} As one consequence of the large and growing
amount of data in individual images (e.g., tens of megabytes available in
individual electron-phase-contrast negatives), formal techniques of physical
inference (like the theory of accessible state probabilities used in
statistical mechanics) can facilitate more rigorous and quantitative study.
In this paper, we outline a strategy for taking steps in this regard. With
Bayes' theorem as our prescription for physical inference from new data,\cite
{Grandy} we first tackle the problem of extracting darkfield information on
periodic structures buried in an otherwise aperiodic field. This problem is
common to electron-phase-contrast and air-based scanning-tunneling images,
even of purely crystalline fields, because specimen preparation and system
instabilities, respectively, often give rise to a superposed aperiodic
background. Second, we discuss how the strategy may be extended to treat a
wider class of problems as well.

\section{Theory}

Mathematically, suppose we are given an experimental image $z[x,y]$ whose
Fourier transform is $\widetilde{{\bf Z}}[u,v]$, and that the objective is
to construct a darkfield image of selected regions of $u$,$v$ space
associated with one or more peaks in the power spectrum $Z^2[u,v]$. Although
we will neglect contributions to the darkfield image from regions of $u$,$v$
space outside of those selected, in the ``peak regions'' we must separate
the $\widetilde{{\bf Z}}[u,v]$ coefficients into peak $\widetilde{{\bf P}}$
and background $\widetilde{{\bf B}}$ components such that $\widetilde{{\bf P}%
}[u,v]\equiv \widetilde{{\bf Z}}[u,v]-\widetilde{{\bf B}}[u,v]$. The
``darkfield image'' is then just the inverse transform of $\widetilde{{\bf P}%
}$, namely $p[x,y]=z[x,y]-b[x,y]$. Here $\widetilde{{\bf bold}}$ print
denotes a complex number. We have assumed also that the starting image is
real, and that $\widetilde{{\bf P}}$ and $\widetilde{{\bf B}}$ as chosen
retain the conjugate symmetry of $\widetilde{{\bf Z}}$ (i.e. $\widetilde{%
{\bf Z}}[-u,-v]=\widetilde{{\bf Z}}^{*}[u,v]$) so that their inverse
transforms are real as well.

In order to separate $\widetilde{{\bf Z}}$ into peak and background
components, information on the nature of the background is crucial.
Previously proposed methods make {\em ad hoc} assumptions about $\widetilde{%
{\bf B}}[u,v]$. For example, the traditional ``window method'' assumes that $%
\widetilde{{\bf B}}[u,v]$ is zero inside the peak regions. The usefulness by
comparison of taking explicit account of estimates for the background
amplitude beneath peaks was recently demonstrated by O'Keefe and Sattler,
but only {\em ad hoc} assumptions (such as a uniformly random distribution)
were proposed for the assignment of background phases.\cite{OKeefe} If,
however, information on background phase is taken into account
systematically, noise in the calculations can be reduced further, and the
background-subtraction recipe can serve as a basis for a plethora of
strategies for direct physical inference.

Specifically, if for any value of $[u,v]$ we consider calculation of the
posterior probability $p(\beta \mid \widetilde{{\bf Z}},B)d\beta $ of the
background phase $\beta $, given the measured coefficient $\widetilde{{\bf Z}%
}{\bf \equiv }Ze^{i\varphi }$ and in addition a prior information value for
the background amplitude $B$, then Bayes' theorem gives

\begin{equation}
p(\beta \mid Z,\varphi ,B)d\beta =p(\beta \mid Z,B)d\beta \frac{p(\varphi
\mid Z,B,\beta )}{p(\varphi \mid Z,B)}\text{.}  \label{Bayes}
\end{equation}
Equation (\ref{Bayes}) tells us how to modify our estimate of the background
phase $\beta $ in light of data on the phase $\varphi $ of $\widetilde{{\bf Z%
}}=\widetilde{{\bf P}}+\widetilde{{\bf B}}$ at that frequency. In notational
terms, $p(\beta \mid Z,B)d\beta $ is the {\em prior\ probability} of $\beta $
given $Z$ and $B$; $p(\varphi \mid Z,B,\beta )d\varphi $ is the {\em %
likelihood function} (or sampling distribution) probability that $\varphi $
will lie between $\varphi $ and $\varphi +d\varphi $ for given values of Z,
B, and $\beta $; and $p(\varphi \mid Z,B)d\varphi $ is the {\em a priori
probability} of $\varphi $ given $Z$ and $B$.

The prior $p(\varphi \mid Z,B)d\varphi $ will of course have no $\beta $
dependence, since it can be considered an integral over all $\beta $.
Determination of such $\beta $-independent terms can be considered academic
here, since $p(\beta \mid Z,B)d\beta $ obeys the normalization condition

\begin{equation}
\int_0^{2\pi }p(\beta \mid Z,B)d\beta =1\text{.}  \label{Norm}
\end{equation}
The $\beta $ dependence of $p(\beta \mid Z,B)d\beta $, one of the
fundamental priors, will be held for discussion below. the $\beta $
dependence of $p(\varphi \mid Z,B,\beta )d\varphi $, on the other hand, is
less fundamental, and is complicated by the fact that $\widetilde{{\bf Z}}$
is a composite of two physically distinct quantities ($\widetilde{{\bf P}}$
and $\widetilde{{\bf B}}$). Fortunately, we can replace this $\varphi $
probability with the product of a probability for $\alpha $, the peak phase
angle, and a geometric term. This is done by writing the peak phase angle $%
\alpha $ in terms of $Z$, $\varphi $, $B$, and $\beta $, and then changing
differential variables in the normalization equation for $p(\alpha \mid
Z,B,\beta )d\alpha $, to get

\begin{equation}
p(\varphi \mid Z,B,\beta )d\varphi =p(\alpha \mid Z,B,\beta )\left| \frac{%
\partial \alpha }{\partial \varphi }\right| _{Z,B,\beta }d\varphi \text{.}
\label{Presto}
\end{equation}
Here the second factor on the right-hand side is the absolute value of the
Jacobian (functional determinant) for the variable change.

To further eliminate $\widetilde{{\bf Z}}$ dependences, note from Bayes'
theorem that

\begin{equation}
p(\alpha \mid Z,B,\beta )d\alpha =p(\alpha \mid B,\beta )d\alpha \frac{%
p(Z\mid \alpha ,B,\beta )}{p(Z\mid B,\beta )}\text{.}  \label{Bayes2}
\end{equation}
As above, a variable change (here from $Z$ to $P$) then allows us to write

\begin{equation}
p(Z\mid \alpha ,B,\beta )dZ=p(P\mid \alpha ,B,\beta )\left| \frac{\partial P%
}{\partial Z}\right| _{\alpha ,B,\beta }dZ\text{.}  \label{Change-O}
\end{equation}
The partials in (\ref{Presto}) and (\ref{Change-O}) are easily calculated,
and hence these expressions allow us to evaluate the probability $p(\beta
\mid Z,\varphi ,B)d\beta $ for {\em any} assignment of the peak and
background prior probabilities $p(\beta \mid Z,B)d\beta $, $p(\alpha \mid
B,\beta )d\alpha $, $p(P\mid \alpha ,B,\beta )dP$.

To evaluate the partials, first note from Fig. 1 that $Z$ can be written in
terms of $P$, $\alpha $, $B$, and $\beta $ as

\begin{equation}
Z=\left\{ \left( P\cos \left[ \alpha \right] +B\cos \left[ \beta \right]
\right) ^2+\left( P\sin \left[ \alpha \right] +B\sin \left[ \beta \right]
\right) ^2\right\} ^{\frac 12}\text{,}  \label{Geom}
\end{equation}
and that $\alpha $ can be related to the variables $Z$, $\varphi $, $B$, and 
$\beta $ via the expressions

\begin{eqnarray}
\varphi -\alpha &=&\arctan \left[ \frac{B\sin \left[ \beta -\varphi \right] 
}{B\cos \left[ \beta -\varphi \right] -Z}\right]  \label{Trig} \\
&=&\arccos \left[ \frac{B\cos \left[ \beta -\varphi \right] -Z}{B^2-2ZB+Z^2}%
\right] \text{.}  \nonumber
\end{eqnarray}
At this point, it is convenient to introduce the following notation for
signed angles in the triangle formed by $\widetilde{{\bf Z}}$, $\widetilde{%
{\bf P}}$, and $\widetilde{{\bf B}}$: $\Theta _B\equiv \varphi -\alpha $, $%
\Theta _P\equiv \beta -\varphi $, and $\Theta _Z\equiv \alpha -\beta -\pi
\times $sgn$\left[ \alpha -\beta \right] $, where sgn$\left[ x\right] \equiv 
$\{$+1$ for $x\geq 0$, $-1$ for $x<0$\}. Note also the mathematical
equivalence of $\Theta _Z$ and $\Theta _B$ under exchange of $\widetilde{%
{\bf Z}}$ and $\widetilde{{\bf B}}$.

By taking the derivative of (\ref{Geom}) with respect to P at constant $%
\alpha $, $B$, and $\beta $, it is easy to show that $\left( \partial
Z/\partial P\right) _{\alpha ,B,\beta }=\widetilde{{\bf Z}}\bullet 
\widetilde{{\bf P}}/ZP=\cos \left[ \Theta _B\right] $, and hence $\left|
\partial P/\partial Z\right| _{\alpha ,B,\beta }=\left| \sec \left[ \Theta
_B\right] \right| $. By differentiating (\ref{Trig}) with respect to $%
\varphi $ at constant $Z$, $B$, and $\beta $, it likewise follows that

\begin{eqnarray}
\left| \frac{\partial \alpha }{\partial \varphi }\right| _{Z,B,\beta }
&=&\left| \frac{1-(B/Z)\cos \left[ \beta -\varphi \right] }{%
(B/Z)^2-2(B/Z)\cos \left[ \beta -\varphi \right] +1}\right|  \label{AtLast}
\\
&=&\frac{\left| \cos \left[ \Theta _B\right] \right| }{P/B}\text{.} 
\nonumber
\end{eqnarray}
Note, therefore, that the product of the two partials is simply $B/P$. These
facts in hand, we can turn to assignment for the prior probabilities $%
p(\beta \mid Z,B)d\beta $, $p(\alpha \mid B,\beta )d\alpha $, and $p(P\mid
\alpha ,B,\beta )dP$ mentioned above.

For the frequent times when prior information singles out no preferred
direction, uniform (isotropic) priors for both $\beta $ and $\alpha $ [i.e., 
$p(\beta \mid Z,B)=p(\alpha \mid B,\beta )=1/2\pi $] will be appropriate. In
any case, these priors do not enter into other potentially $\beta $%
-dependent terms in Eqs. (\ref{Bayes})-(\ref{Change-O}). Hence, should we
assign them an explicit $\beta $-dependence it will be factored into $%
p(\beta \mid Z,\varphi ,\beta )$ directly.

The most complicated prior is $p(P\mid \alpha ,B,\beta )$, because it cannot
be dismissed by isotropy, and because it enters also into the only term
whose $\beta $-dependence remains undiscussed, namely $p(Z\mid B,\beta )$ in
Eq.(\ref{Bayes2}). We point out here two choices of potential interest. A
prior for $P$ which (i) appears simple to implement, (ii) results in
correlations as expected between the direction of $B$ and $Z$ when $B\approx
Z$, but (iii) has unclear physical meaning at best is that which makes $%
p(\alpha \mid Z,B,\beta )$ in Eq. (\ref{Presto}) a constant. The probability
distribution for $\beta $ resulting from such a prior is plotted in Fig. 2.
To generate pseudorandom values for $\beta $ with this distribution, one can
simply choose values of $\Theta _Z$ (hence $\alpha $) uniformly distributed
in its allowed range for given $Z/B$, and then calculate $\beta -\varphi
=\Theta _P$ from $\Theta _Z$ allowing for the fact that $\Theta _P$ is a
double-valued function of $\Theta _Z$ with 50\% of its values on each branch
when $Z<B$. Numerical simulations in our laboratory have verified that a
pseudorandom-number generator so constructed reproduces the distribution
shown in Fig. 2.

The physically (if not computationally) most helpful assignment for $p(P\mid
\alpha ,B,\beta )$ is likely to be the uniform ($\beta $-independent)
assignment. If we view $P$ not as a scale factor but as the magnitude of a
vector translation whose length is scaled to the value of $B$, then $p(P\mid
\alpha ,B,\beta )=const=1/P_{\max }$ for P$_{\max }\gg Z_{\max }$ may be
considered the correct transformation-invariant prior probability in the
absence of further information.\cite{Jaynes} In this case,

\begin{equation}
p(Z\mid B,\beta )=\frac 1{P_{\max }}\int_0^{2\pi }\left| \sec \left[ \Theta
_B\right] \right| d\Theta _Z\text{.}  \label{Physical}
\end{equation}
The argument in this integral is double-valued and must therefore be handled
carefully for $\Theta _B<\Theta _Z$, but it can nevertheless be written as a
function of $Z$, $B$, and $\Theta _Z$ only. Hence the integral itself is
independent of $\beta $. Given this uniform prior for $P$, and isotropic
priors for $\alpha $ and $\beta $, Eqs. (\ref{Bayes})-(\ref{Change-O})
therefore give a $\beta $-dependence for $p(\beta \mid Z,\varphi ,B)$
proportional to the product of partials, and inversely proportional to $P$.
This distribution is plotted for various $Z/B$ in Fig. 3. Numerical
simulations in our laboratory have confirmed that $P$ vectors with a uniform
distribution of lengths and orientation angles yield $\beta $-values with
the distribution shown.

\section{Summary}

The Bayesian phase model here does two things. First of all, it takes the
image-processing recipe for ``background subtraction'' proposed by O'Keefe
and Sattler and creates a recipe for physical inference by replacing {\em ad
hoc} assignments of background phase with an assignment of phases based on
physical assumption. Second, it systematically minimizes noise in darkfield
images associated with crystallite fine structure (e.g., boundaries), since
the information on fine structure is found in outlying regions of the Bragg
peaks where $Z/B$ is near $1$. These are the regions where errors associated
with an {\em ad hoc} phase assignment will have their greatest effect.
Because the image itself, not the imaging process, constitutes the object of
the physics in this analysis, it can be applied to images inferred via
Bayesian statistical analysis of the imaging process\cite{Kirkland} as well
as to raw experimental data.

Strategies for obtaining prior information on the amplitude of $B$ in these
applications begin with simple extrapolation of an azimuthally symmetric
``cloud'' of aperiodic contrast in Fourier space.\cite{Fraundorf}
Applications of Fourier darkfield analysis which are now limited by
resolution and signal-to-noise ratio include the study of isotopically
anomalous nm-sized diamonds in meteorites,\cite{Fraundorf2} and the study of
objects showing space-group disallowed symmetries due to twinning and
quasicrystallinity. The algorithms here facilitate removal of aperiodic
background from selected frequency-space regions in images of these
structures. Some applications, such as the study of icosahedral structures
in periodic array,\cite{Levine} may require removal of periodic background.
These involve prior information which goes beyond the scope of the result,
but not the formalism, discussed here.

The formalism is also not limited to prior information on background
amplitude. Prior information on the variance in background amplitude (here
assumed to be zero) is a logical addition. When it can be made a practical
addition remains to be seen. The formalism should also allow us to estimate
the strength of periodicities (analogous to diffracted beam intensities)
from point to point in the field. This is something that diffraction
darkfield does which Fourier darkfield does less directly. However, the
strategy posed here for physical inference from images allows this and other
questions to be asked mathematically. As another example, diffraction
darkfield imaging does not allow us to distinguish between crystals with the
same periodicity but a different periodicity phase. Atomic-resolution images
with data on periodicity phase and amplitude contain the needed data, and
the strategy proposed here suggests a formalism for posing this question as
well.

\begin{figure}[tbp]
\caption{Given complex Fourier coefficients obeying $Ze^{i\varphi
}=Pe^{i\alpha }+Be^{i\beta }$, this schematic illustrates the geometric
relationship between peak phase angle $\alpha $, and background phase angle $%
\beta $, when $Z$, $\varphi $, and $B$ are specified.}
\label{Fig1}
\end{figure}

\begin{figure}[tbp]
\caption{A plot of $p(\beta \mid Z,\varphi ,B)$ in probability per radian
interval vs. $\beta -\varphi $ in radians for the computationally convenient
case when $p(\beta \mid Z,B)=1/2\pi $, and $p(\alpha \mid Z,B,\beta )$ is
independent of $\beta $. Note the strong bias toward $\beta \approx \varphi $
as $Z/B$ approaches $1$, and the cusps for $Z/B<1$. An algorithm for
generating $\beta $ values with this distribution is described in the text.
Notation for $Z/B$: solid line, $\frac 18$; plus, $\frac 12$; cross, $2$;
diamond, $8$.}
\label{Fig2}
\end{figure}
\begin{figure}[tbp]
\caption{A plot of $p(\beta \mid Z,\varphi ,B)$ in probability per radian
interval vs. $\beta -\varphi $ in radians for the physically realistic case
when $p(\beta \mid Z,B)=p(\alpha \mid B,\beta )=1/2\pi $, and $p(P\mid
\alpha ,B,\beta )=const=1/P_{\max }$, where $P_{\max }\gg Z_{\max }$. Note
the bias toward $\beta \approx \varphi $ as $Z/B$ approaches $1$. The bias
will weaken with increasing variance in $B$ when nonzero variances are
included as prior information in the calculation. Notation for $Z/B$: solid
line, $\{512,\frac 1{512}\}$; plus, $\{8,\frac 18\}$; cross, $\{2,\frac 12\}$%
; diamond, $\{2^{1/3},2^{-1/3}\}$.}
\label{Fig3}
\end{figure}

\end{document}